\documentclass[twocolumn,twocolappendix]{aastex701}

\usepackage{amsmath}
\usepackage{longtable}
\usepackage{{booktabs}}
\newcommand{\gwsn}{\texttt{GWSkyNet}}

\newcommand{\gwsnm}{\texttt{GWSkyNet-Multi}}
\newcommand{\gwsnmm}{\texttt{GWSkyNet-Multi~II}}
\newcommand{\gwsnmg}{\texttt{GWSkyNet-MassGap}}

\shorttitle{GWSkyNet-MassGap}
\shortauthors{Raza et al.}

\begin{document}

\title{Training a neural network to rapidly identify candidate \\ gravitational-wave events in the lower mass gap}

\author[orcid=0000-0002-8549-9124,gname=Nayyer,sname=Raza]{Nayyer Raza}
\affiliation{Department of Physics, McGill University, 
3600 rue University, Montr\'{e}al, QC H3A2T8, Canada}
\affiliation{Trottier Space Institute at McGill, 
3550 rue University, Montr\'{e}al, QC H3A2A7, Canada}
\email[show]{nayyer.raza@mail.mcgill.ca}

\author[orcid=0009-0009-6826-4559,gname='Man Leong',sname=Chan]{Man Leong Chan}
\affiliation{Department of Physics and Astronomy, University of British Columbia, 
Vancouver, BC V6T1Z1, Canada}
\email{mervync@phas.ubc.ca}

\author[orcid=0000-0001-6803-2138,gname=Daryl,sname=Haggard]{Daryl Haggard}
\affiliation{Department of Physics, McGill University, 
3600 rue University, Montr\'{e}al, QC H3A2T8, Canada}
\affiliation{Trottier Space Institute at McGill, 
3550 rue University, Montr\'{e}al, QC H3A2A7, Canada}
\email{daryl.haggard@mcgill.ca}

\author[orcid=0000-0003-2242-0244,gname=Ashish,sname=Mahabal]{Ashish Mahabal}
\affiliation{Division of Physics, Mathematics and Astronomy, California Institute of Technology, Pasadena, CA 91125, USA}
\affiliation{Center for Data Driven Discovery, California Institute of Technology, Pasadena, CA 91125, USA}
\email{aam@astro.caltech.edu}

\author[orcid=0000-0003-0316-1355,gname=Jess,sname=McIver]{Jess McIver}
\affiliation{Department of Physics and Astronomy, University of British Columbia, Vancouver, BC V6T1Z1, Canada}
\email{jess.mciver@ubc.ca}

\author[orcid=0000-0001-9290-8603,gname=Audrey,sname=Durand]{Audrey Durand}
\affiliation{Department of Computer Science and Software Engineering, Universit\'{e} Laval, Qu\'{e}bec, QC G1V0A6, Canada}
\affiliation{Canada CIFAR AI Chair, Mila, Montr\'{e}al, QC H2S3H1, Canada}
\email{audrey.durand@ift.ulaval.ca}

\author[orcid=0009-0002-0606-7516,gname=Alexandre,sname=Larouche]{Alexandre Larouche}
\affiliation{Department of Computer Science and Software Engineering, Universit\'{e} Laval, Qu\'{e}bec, QC G1V0A6, Canada}
\email{alexandre.larouche.7@ulaval.ca}

\author[orcid=0000-0003-3832-8856,gname=Hadi,sname=Moazen]{Hadi Moazen}
\affiliation{Department of Computer Science and Software Engineering, Universit\'{e} Laval, Qu\'{e}bec, QC G1V0A6, Canada}
\email{hadi.moazen.1@ulaval.ca}

\correspondingauthor{Nayyer Raza}

\begin{abstract}

The physics governing the boundary between the most massive neutron stars (NSs) and the least massive black holes (BHs) is currently uncertain, but could potentially be constrained with new observations. While NSs have been observed with masses up to $\sim2~M_{\odot}$, there is a dearth of electromagnetic observations of compact objects in the $\sim2-5~M_{\odot}$ range, known as the lower mass gap. Recent observations of gravitational-wave (GW) signals from binary mergers detected by the LIGO-Virgo-KAGRA (LVK) collaboration indicate that this gap is likely not empty. Rapidly distinguishing whether a candidate GW event has components in this purported mass gap can indicate the likelihood of a detectable electromagnetic counterpart, and thus inform decisions for follow-up observations. In this work we train a neural network model, \gwsnmg, that simultaneously predicts the probability that a candidate merger has a component in the lower mass gap ($P_{\mathrm{MassGap}}$) and the probability that it involves a NS ($P_{\mathrm{NS}}$). We find that the model is able to infer information about the source chirp mass to predict $P_{\mathrm{MassGap}}$ and $P_{\mathrm{NS}}$, leading to correct predictions for high-mass mergers with $\mathcal{M}_c\gtrsim15~M_{\odot}$, but less accurate predictions for lower-mass systems which require knowledge of the binary mass ratio to break the mass degeneracy. For candidate events in the first part of LVK's fourth observing run (O4a), the model has a mean prediction error of 9\% for $P_{\mathrm{MassGap}}$ and 6\% for $P_{\mathrm{NS}}$. The model could be further developed to rapidly predict the source chirp mass for candidate events in future observing runs.

\end{abstract}

\keywords{\uat{Gravitational wave astronomy}{675} --- \uat{Gravitational wave sources}{677} --- \uat{Neural networks}{1933}}

\section{Introduction} \label{sec:intro}

The LIGO-Virgo-KAGRA (LVK) gravitational-wave (GW) observatories \citep{Acernese2015,Aasi2015,Akutsu2021} have reported 218 significant binary merger candidates with astrophysical probability $p_{\mathrm{astro}} > 0.5$ up to the end of the first part of the fourth observing run, O4a \citep{Abbott2019_gwtc1,Abbott2023_gwtc3,Abbott2024_gwtc2.1,Abac2025_gwtc4}. These events include binary black hole (BBH) mergers, neutron star-black hole (NSBH) mergers, and binary neutron star (BNS) mergers.

However, not all merger events can be confidently classified into one of the three aforementioned categories, because the boundary for the maximum neutron star (NS) mass is dependent on its dense matter equation of state (EOS) and the precise value is not known. Observations of the most massive known pulsar to date, PSR J0740+6620, suggest a lower limit for the maximum NS mass of $\sim 2 ~M_{\odot}$ \citep{Fonseca2021,Legred2021}, while theoretical considerations constrain the upper limit to be $\lesssim 3~M_{\odot}$ \citep[e.g.,][]{Kalogera1996}. At the same time, the transition between the most massive known NSs and the least massive black holes (BHs) is currently unclear: analysis of the stellar-mass BH mass distribution from dynamical measurements of X-ray binaries has suggested the possible existence of a lower mass ``gap'' in the $\sim 3-5 ~M_{\odot}$ range (\citealt{Bailyn1998, Ozel2010, Farr2011} --- to be distinguished from the upper mass gap for BHs due to pair-instability supernovae, which does not involve NSs; see, e.g., \citealt{Woosley2021} and references therein). However, recent GW detections of binary merger events have now started to populate this region and provide evidence that it is not empty, with events GW190814 \citep{Abbott2020_GW190814}, GW200115 \citep{Abbott2021ApJ_GW200115}, and GW230529 \citep{Abac2024_GW230529} all having a high probability that at least one of their binary components lies in the lower mass ``gap''. Detection of compact objects with masses in this range of $\sim 2-5 ~M_{\odot}$ thus represent an interesting class of events that can provide insights into the boundary between NSs and BHs.

Merger events that involve a NS are prime candidates for having associated electromagnetic (EM) bright emission. BNS mergers are known to produce kilonovae and short gamma-ray bursts (GRBs), as was observed for GW170817 \citep{Abbott2017_gw170817, Abbott2017_multimessenger, Abbott2017_grb170817a}. NSBH mergers can also produce associated electromagnetic counterparts (e.g., kilonovae and GRBs) if the NS is tidally disrupted by the BH before merger. This is more likely to occur if the mass ratio of the binary $q = m_1/m_2$ is low, the compactness of the NS is small, and the spin of the BH is high \citep[see, e.g., a review of NSBH binaries in][]{Kyutoku2021}.

In particular, the low mass ratio condition implies that NSBH binaries that have the primary component (BH) in the mass gap have a higher likelihood of producing an EM counterpart, as opposed to NSBH binaries that have the primary component mass beyond the mass gap ($m_1 \gtrsim 5~M_{\odot}$). Conversely, for a binary that has a component in the mass gap, which could potentially either be a NS or BH, the EM bright potential is determined by the probability that at least one of the binary components is a NS. Simultaneously estimating the probability that a binary has a component in the mass gap and a component that is a NS gives the most complete information regarding the likelihood of the source having an EM counterpart, and thus whether it is a high-interest candidate for potential follow-up observations.

In this work we expand on the \gwsn \ \citep{Cabero2020,Chan2024} and \gwsnm \ \citep{Abbott2022,Raza2025} machine learning models developed for rapid classification of candidate GW events. We train a neural network model to simultaneously predict the probability that the event has a binary component in the lower mass gap and the probability that it has a component that is a NS. These predictions build upon the mutually exclusive classifications of \gwsn \ (glitch vs astrophysical) and \gwsnm \ (glitch vs BNS vs NSBH vs BBH), and thus provide additional information about the candidate event properties within our framework of models to aid observation decisions in real-time, complementing the source property annotations released by the LVK in public alerts to the follow-up community \citep{Chaudhary2024}.

The new model, which we call \gwsnmg, makes use of the same inputs as \gwsnm: the predictions are based on the rapidly computed \textsc{Bayestar} \citep{Singer2016} source localization information for candidate merger events detected by the LVK, as provided to the community through the low-latency public alerts and available on the Gravitational-Wave Candidate Event Database (GraceDB)\footnote{\url{https://gracedb.ligo.org}}. However, \gwsnmg \ is trained on a larger and more astrophysically robust set of simulated merger events. While \gwsnm \ was trained on data that was generated from discrete binary source populations that did not cover the ambiguous mass gap events, in this work we model the sources from a global distribution of all compact binary mergers. In particular, instead of assuming mass cut-off values for binary classification of component events (NS vs BH), we take a more nuanced approach to reflect the known  uncertainty in our labeling of these events: we calculate and train the model directly on the \textit{probabilities} of the event belonging to each category. This approach avoids assuming sharp cut-off boundaries for the component masses to classify events, which is necessary since the nature of events in the lower mass gap can be ambiguous.

We note that the LVK also provide these target properties in low-latency as part of the public alert contents: \texttt{HasNS} and \texttt{HasMassGap} \citep[in addition to \texttt{HasRemnant};][]{Chaudhary2024, Chatterjee2020}. These properties are also directly predicted by supervised machine learning models that take as input the parameters from the online best-matched template: component masses, component spins, and network signal-to-noise ratio (SNR). \texttt{HasNS} uses a nearest-neighbor classifier algorithm, marginalizing over several different NS equation-of-state (EOS), and \texttt{HasMassGap} uses a random-forest classifier algorithm, indicating events as MassGap which have $3~M_{\odot} < m < 5~M_{\odot}$ \citep{Chaudhary2024}. However, these five input parameter values are not part of the data released in O4 public alerts\footnote{\url{https://emfollow.docs.ligo.org/userguide/}}. Our work differs in that it uses an entirely different parameter space of inputs to predict the MassGap and NS properties: data from the \textsc{Bayestar} localization maps that are released in public alerts. Thus it offers a complementary determination of these source properties using a model that is also transparent in terms of its inputs and fully open-source to the community.

This paper is organized as follows: in Section \ref{sec:methods} we describe how we generate our simulated data set, determine the source probabilities of mass gap and NS, and train the model to predict these values. In Section \ref{sec:results} we evaluate the model's performance on test set data, analyze the results to find trends in the model predictions, and compare the model's predictions for events in O4a. In Section \ref{sec:model_data_limitations} we discuss some limitations of the model and training data. In the final section we summarize and offer concluding remarks.

\section{Methods} \label{sec:methods}

\subsection{Simulations of astrophysical merger events}

We generate our set of simulated binary merger events by sampling event properties from astrophysically motivated source distributions, modeling the corresponding GW signal waveforms, injecting these waveforms into realistic estimates of the O4 detector network sensitivities, and recovering the localization properties of events that pass our detection threshold.

We draw our component mass samples ($m_1, m_2$) from the \textsc{Power Law + Dip + Break} (PDB) parametrized mass distribution model \citep{Fishbach2020,Farah2022, Abbott2023_gwtc3_population} with random (independent) pairing for the components in the range $m_1,m_2 \in [1,100]~M_{\odot}$, imposing $m_1 \leq m_2$. The PDB model simultaneously fits the entire population of compact binaries without assuming any mass cut-offs for sub-population classification. Additionally, it employs a notch-filter dip to account for a potential lower mass gap at the location of the power law break, where the amplitude of the dip $A$ is a parameter of the model and controls the degree to which mergers in the potential mass gap are suppressed, and the width of the gap is controlled by the location of the lower and upper edges of the mass gap,  $M^{\mathrm{gap}}_{\mathrm{low}}$ and $M^{\mathrm{gap}}_{\mathrm{high}}$. The PDB model we use to generate our data set is the mass distribution model fit to the entire compact binary population in the LVK GWTC-3 catalog and presented as the fiducial population model in \cite{Abbott2023_gwtc3_population}. The PDB model also assumes the spin magnitudes to be uniformly distributed, with a maximum value of $\chi=0.4$ for $m < 2.5 ~M_{\odot}$ and $\chi=1$ for $m \geq 2.5 ~M_{\odot}$. To account for the uncertainty in the population parameter inference, we follow a similar procedure as \cite{Kunnumkai2024} and draw our mass samples from the full hyper-posterior Monte Carlo distribution for the PDB model fit to the events in GWTC-3, using the data available at \cite{LVK2021_gwtc3_data}. Sampling from the full posterior distribution is important since some parameters have broad support over the entire prior range: the median and 90\% confidence interval for the mass gap depth, for example, is $A=0.77^{+0.20}_{-0.47}$. 

With the mass and spin parameters ($m_1, m_2, \chi_1, \chi_2$) drawn from the PDB model describing the intrinsic source parameters, we then draw samples for the extrinsic parameters: the distances are sampled to be uniform in volume, the sky location (right ascension, declination) uniform on sky, and the inclination and polarization angles uniform in orbital orientation. The time of the event is sampled uniformly over a 24-hour period; as we assume a fixed, time-independent detector sensitivity, the time of the event only makes a difference on a 24-hour timescale during which the detectors are sensitive to different parts of the sky due to the Earth's rotation.

For the detector sensitivities we use the O4 noise power spectral density curves estimated by the LVK before the start of O4 and provided for use in simulations \citep{LVKO4PSD2022}. We use the low sensitivity estimates for the LIGO (\texttt{aligo\_O4low.txt}) and Virgo (\texttt{O4\_Virgo\_78.txt}) detectors, as opposed to the high sensitivity estimates, as these more closely match the actual detector sensitivities achieved during O4 \citep{Capote2025}. Similar to the detector duty cycle values used in \cite{Shah2024}, we assume a duty cycle of 70\% for the LIGO Hanford and Livingston detectors, and a lower value of 50\% for the Virgo detector to account for the fact that it was not online during O4a (O4a represents $\sim 30\%$ of the total O4 observing time, and so the Virgo duty cycle is reduced to $\sim 70\%$ of the LIGO value). For our event simulations we then draw samples for the online detector network proportional to their joint up-times, limiting them to multi-detector configurations: $[P_{\mathrm{HL}}, P_{\mathrm{HV}}, P_{\mathrm{LV}}, P_{\mathrm{HLV}}] = [0.35, 0.15, 0.15, 0.35]$. Note that this is not the same as the proportion of events \textit{detected} in each network configuration, which is additionally dependent on the distribution of event sources and the different detector network sensitivities. For our final generated data set of detected events, we find that 37\% are HL events, 9.6\% HV, 9.4\% LV, and 44\% HLV. As was done for \gwsn \ and \gwsnm, we limit the events considered in this work to be multi-detector only as single detector events are not well localized and thus not conducive to follow-up observations.

To simulate the astrophysical event waveforms we use the \texttt{TaylorF2} \citep{Mishra2016} waveform approximant model for events with $m_1 + m_2 < 4 ~M_{\odot}$, and the \texttt{SEOBNRv4-ROM}
\citep{Bohe2017} model for higher masses. These waveform models are chosen to match the ones used by the \textsc{PyCBC} and \textsc{GstLAL} search pipelines to identify events in GWTC-3 \citep{Abbott2023_gwtc3}. To account for potential biases that might be introduced by our choice of waveform models, we also simulate astrophysical events using the \texttt{IMRPhenomD} \citep{Khan2016} model for the entire range of masses considered ($m_1,m_2 \in [1,100]~M_{\odot}$). Similar to the approach taken by the LVK in reporting the fiducial parameter estimation results for a candidate event, in this work we generate and analyze an equally mixed sample of events, so that half the samples are generated with the \texttt{TaylorF2 + SEOBNRv4-ROM} waveforms and half with the \texttt{IMRPhenomD} waveforms. All three of these waveform models are also used for identifying candidate events in GWTC-4.0 \citep{Abac2025_gwtc4_methods}. We use the LALSuite software \citep{lalsuite2018} to generate these waveforms with the sampled source parameters described in the previous paragraph. The three waveform models are limited to aligned-spin systems for computational efficiency (which is necessary to identify events in low-latency), and so we also limit the component spins of our samples to lie parallel to the orbital angular momentum vector: $\chi = [0, 0, \chi_z]$.

Each waveform is then injected into Gaussian noise colored by the simulated O4 detector power spectral density curves, in a 128 second segment sampled at 4096 Hz. We use the \texttt{ligo.skymap}\footnote{\url{https://git.ligo.org/lscsoft/ligo.skymap}} package to simulate running a matched-filter search pipeline to detect the injected signal with Gaussian measurement error and produce the corresponding \textsc{Bayestar} localization map. For this analysis we select a low-frequency cut-off of $f_{\mathrm{low}} = 20~\mathrm{Hz}$ and a high frequency cutoff at the Nyquist frequency of $f_{\mathrm{high}} = 2048~\mathrm{Hz}$. A network signal-to-noise ratio (SNR) threshold of $\rho_{\mathrm{net}} > 7$ is applied to exclude weakly detected signals that are not likely to be identified as astrophysical events by the LVK search pipelines; of the events included in GWTC-3 the lowest SNR event has $\rho_{\mathrm{net}} \simeq 7.2$ \citep[excluding the two outlier events for which the parameter estimation is dominated by potentially unphysical, low-likelihood modes;][]{Abbott2023_gwtc3}. This network SNR threshold also ensures that for the multi-detector events considered in this work, at least one detector has $\rho_{\mathrm{det}} > 4$.

We also impose a maximum distance cutoff of $d_{max} = 2.4 ~\mathrm{Gpc}$ for samples in our data set. This upper limit is chosen based on the sensitivity of the LIGO-Virgo detectors to MassGap and NS events in O4: based on the samples drawn from the PDB model we find that this distance covers $99\%$ of detectable events with $m_2 < 3 ~M_{\odot}$ (i.e., potential NS events) and $\sim 95\%$ of detectable events with $m_2 < 6 ~M_{\odot}$ (potential MassGap events). Here we use the network SNR threshold of $\rho_{\mathrm{net}} > 7$ to define detectable events in O4. Thus virtually all binary mergers detected in O4 with source distances beyond $d\sim 2.4 ~\mathrm{Gpc}$ are expected to be unambiguous BBHs; the maximum distance cutoff ensures that our data set is not overly dominated by these BBH events, while still accurately modeling the BBH distribution at smaller distances. The model training can then focus on this more interesting regime of detectable mergers at smaller distances, as they could potentially be MassGap or NS events.

Drawing our samples from the aforementioned source and detector distributions, we generate a total of $2\times 10^4$ merger events that pass our SNR and distance thresholds, and form the final data set used in this work. The number of events is chosen to balance having enough samples to cover the entire parameter space of events (see, e.g., Figure~\ref{fig:m1_m2_massgap_ns_prob}) against limited computational time to perform the simulations. For each of these events we have an associated \textsc{Bayestar} localization FITS file, mirroring what is released by the LVK for candidate CBC events in O4 public alerts, which we then process to calculate the relevant model inputs for \gwsnmg.

\subsection{Determination of MassGap and NS probabilities}

\begin{figure}
\includegraphics[width=\columnwidth]{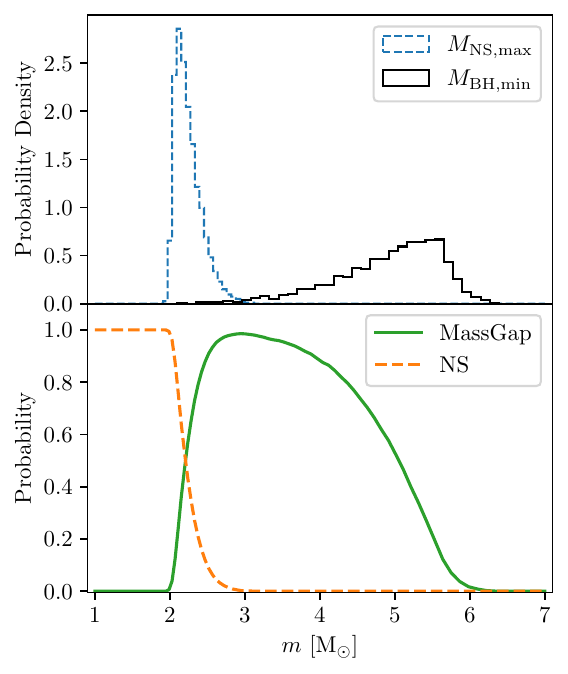}
\caption{Probability that a compact object with mass $m$ is a neutron star (NS) and probability that it is in the lower mass gap (MassGap) (bottom panel). The probabilities are determined from the posterior samples for the maximum NS mass from \cite{Legred2021} and the minimum BH mass from the \textsc{PowerLaw+Peak} model from \cite{Abbott2023_gwtc3_population} (top panel). The probabilities are then used to calculate the combined probability that a binary system has at least one NS component, $P_{\mathrm{NS}}$, and the probability that it has at least one component in the lower mass gap, $P_{\mathrm{MassGap}}$, following Eqs.~\ref{eq:1}-\ref{eq:3}.
\label{fig:NS_BH_MassGap_probs}}
\end{figure}

\begin{figure*}
\includegraphics[width=\textwidth]{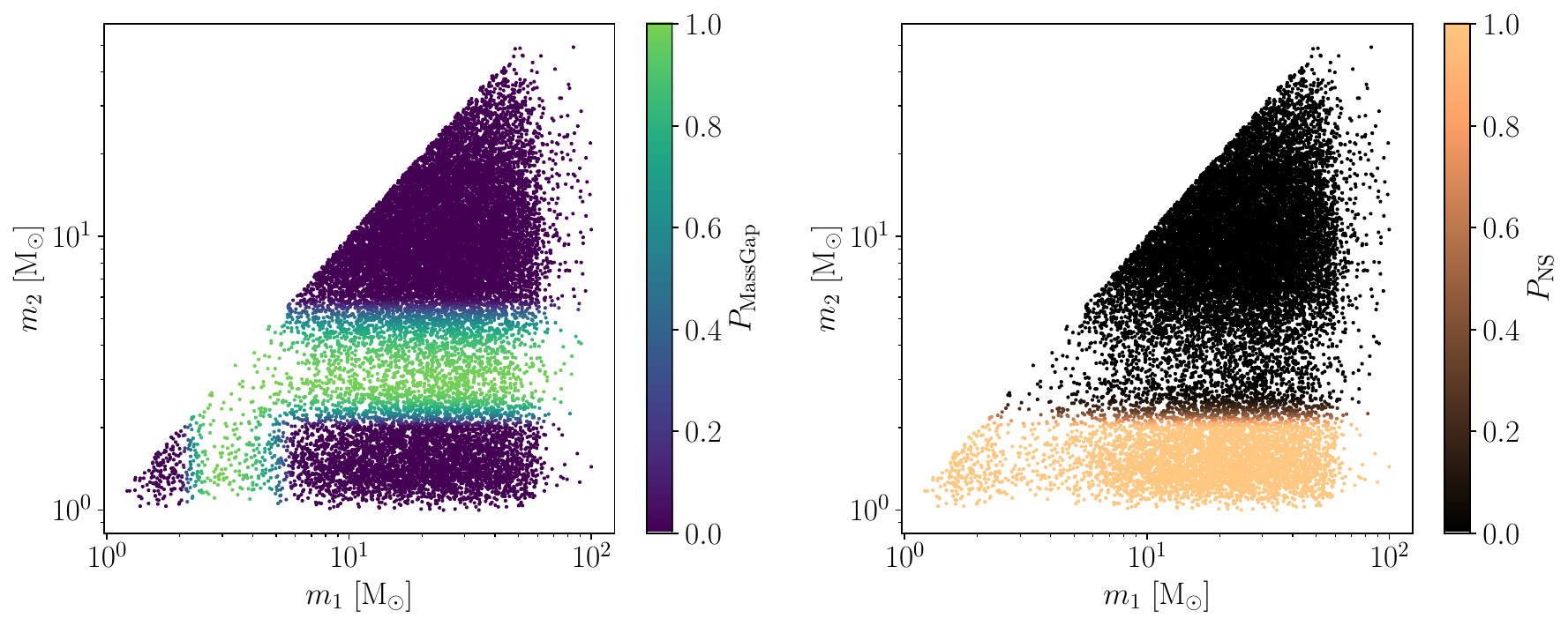}
\caption{Scatter plot showing the component masses ($m_1, m_2$) of all $2 \times 10^4$ events in our generated data set, and their determined MassGap and NS probabilities according to Eqs.~\ref{eq:1}-\ref{eq:3}. The density of points in the mass gap range of $\sim 2.2 - 5.1 ~M_{\odot}$ shows that while a lower fraction of events occur in this range, the mass ``gap'' is not empty. 14\% of events in our data set have $P_{\mathrm{MassGap}} > 0.5$, while 24\% have $P_{\mathrm{NS}} > 0.5$ (note that these fractions are larger than the true astrophysical detection rates because of the maximum distance cut-off we impose to balance the data set). We also see a general drop-off in the number of events with component masses $m_1 \gtrsim 60 ~M_{\odot}$ and $m_2 \gtrsim 40 ~M_{\odot}$, as the PDB model predicts a much smaller merger rate for these high mass systems.
\label{fig:m1_m2_massgap_ns_prob}}
\end{figure*}

The classification of events to train the model in \gwsnm \ was based on a mass threshold value to decide whether each component in a binary merger was a NS or BH. In \gwsnm \ this threshold was chosen to be a conservative upper value of $m = 3 ~M_{\odot}$, while for \gwsnmm \ this was lowered to a more realistic value of $m = 2.5 ~M_{\odot}$ based on an analysis of GW merger events with NSs \citep{Landry2021}. In this work we take a different approach and instead of assuming a single threshold value to label a component in a binary fashion, we follow the methodology in \cite{Essick2020} and marginalize over the range of possible values to determine the probability that the component mass is consistent with that of a NS. To calculate this probability we compare the component mass to the maximum mass estimate for a non-rotating NS, the Tolman–Oppenheimer–Volkoff (TOV) mass $M_{\mathrm{max,TOV}}$. We follow the approach used in \cite{Abbott2023_gwtc3_population} and draw the hyper-posterior samples for $M_{\mathrm{max,TOV}}$ from an EOS inference analysis which jointly fits data from pulsar timing, gravitational-wave and X-ray observations of NSs \citep{Legred2021}. These are shown in Figure~\ref{fig:NS_BH_MassGap_probs}, and we consider them to be the values for the maximum NS mass, $M_{\mathrm{NS,max}}$. We then calculate the probability of a component with mass $m$ to be a NS as $P_{\mathrm{NS}}(m) = P(m < M_{\mathrm{NS,max}})$, marginalizing over the posterior distribution of all $M_{\mathrm{NS,max}}$ values, with the results shown in the bottom panel of Figure~\ref{fig:NS_BH_MassGap_probs}. Finally, the probability that a binary with masses $(m_1, ~m_2)$ has \textit{at least one} NS component is:
\begin{multline}\label{eq:1}
    P_{\mathrm{NS}}(m_1,m_2) = 1 - [P(m_1 > M_{\mathrm{NS,max}}) \\
    \times P(m_2 > M_{\mathrm{NS,max}})] \ \qquad
\end{multline}
The calculated binary $P_{\mathrm{NS}}$ for all $(m_1, ~m_2)$ pairs in the $2\times 10^4$ events in our data set are shown in the right panel of Figure~\ref{fig:m1_m2_massgap_ns_prob}.

To determine the probability of a binary having at least one component in the lower mass gap, $P_{\mathrm{MassGap}}$, we need both a lower and upper bound on the masses that span this mass gap range. Since the mass gap signifies the transition between NSs and BHs, the natural choices for choosing the lower  and upper limits are then the maximum NS mass $M_{\mathrm{NS,max}}$ and the minimum BH mass $M_{\mathrm{BH,min}}$, respectively. To calculate the minimum BH mass, we use the $m_{min}$ parameter of the \textsc{Power Law + Peak} (PP) mass distribution model used to fit the population of BBHs in GWTC-3 \citep{Abbott2023_gwtc3_population}. We draw the hyper-posterior samples for this fit parameter from the data release \citep{LVK2021_gwtc3_data}, and label them as $M_{\mathrm{BH,min}}$, as shown in the top panel of Figure~\ref{fig:NS_BH_MassGap_probs}. We then calculate the probability of a component with mass $m$ to be in the mass gap as 
\begin{equation}
    P_{\mathrm{MassGap}}(m) = P(m > M_{\mathrm{NS,max}}) \times P(m < M_{\mathrm{BH,min}}),
\end{equation}
this time marginalizing over the posterior distribution of all $M_{\mathrm{NS,max}}$ and $M_{\mathrm{BH,min}}$values. The calculated values are shown in the bottom panel of Figure~\ref{fig:NS_BH_MassGap_probs}. Finally, the probability that a binary with masses $(m_1, ~m_2)$ has \textit{at least one} component in the mass gap is:
\begin{multline}\label{eq:3}
    P_{\mathrm{MassGap}}(m_1,m_2) = P_{\mathrm{MassGap}}(m_1) + P_{\mathrm{MassGap}}(m_2) \\ 
    - [P_{\mathrm{MassGap}}(m_1) \times P_{\mathrm{MassGap}}(m_2)]
\end{multline}

The calculated binary $P_{\mathrm{MassGap}}$ for all $(m_1, ~m_2)$ pairs in the $2\times 10^4$ events in our data set are shown in the left panel of Figure~\ref{fig:m1_m2_massgap_ns_prob}.

\subsection{Model architecture and training}

Of the $2\times 10^4$ events in our data set, we split the data so that 50\% is used for training the model and 50\% is reserved for testing its predictions after the model has been trained. We choose a relatively high fraction of data to use for our test set as we aim to study these event predictions to find robust trends in what the model might have learned. Furthermore, we find that using more than $\sim 10^4$ events to train the model does not lead to any improvement in its performance, and so we limit the training set to this size. 

The inputs to the \gwsnmg \ model are values extracted from the \textsc{Bayestar} sky localization map: (1) 90\% sky localization area, (2) 90\% volume localization, (3) mean estimated distance, (4) standard deviation of estimated distance, (5) Log Bayes signal-versus-noise factor (Log BSN), (6) Log Bayes coherence-versus-incoherence factor (Log BCI), and (7) the detector network observing at the time of the event. These inputs are the same as in \gwsnmm, with one minor difference: the Log BSN factor was clipped at a maximum value of 100 for input to \gwsnmm, whereas in this work the maximum value is increased and values clipped beyond $\mathrm{Log ~BSN} > 200$ (the new maximum value chosen so that 95\% of events in our data have Log BSN below this value). This allows the model to learn from a slightly larger range of values for Log BSN, while still balancing sensitivity to small differences in the value (we also empirically test this threshold value during model training and find that the model performance does indeed marginally decrease when the maximum Log BSN value is increased beyond 200).

The model architecture is a neural network similar to that for \gwsnmm: a sequence of dense (fully-connected) layers with each neuron in the layer having a rectified linear unit activation function. However, the number of hidden layers and number of neurons in each layer are modified to best predict the new target variables, $P_{\mathrm{MassGap}}$ and $P_{\mathrm{NS}}$. Thus the output layer in \gwsnmg \ is a dense layer of two neurons with a \textit{sigmoid} activation function for each to simultaneously output the MassGap and NS probabilities. The number of hidden layers, number of neurons in each layer, training batch size, learning rate, and patience for early-stopping are hyper-parameters that are tuned using a mix of automatic hyperparameter optimization with the Optuna framework \citep{Akiba2019} and manual fine-tuning. During model training we further split the training data so that 10\% (1000 events) are reserved for model validation, which allows us to evaluate the model's performance on this independent set at the end of every epoch and prevent model over-fitting to the training data. In all cases, the hyper-parameters are tuned to achieve the best mean squared error loss on the validation data set. The best resulting model is trained with a batch size of 128, a learning rate of 0.001, an early-stopping patience of 200 epochs (with a maximum number of 5000 epochs allowed for training), and has four hidden dense layers with 32 neurons in each layer. Compared to \gwsnmm, which has two hidden dense layers with eight neurons in each  layer, \gwsnmg \ has twice as many hidden layers and four times the number of neurons in each layer (the model is deeper and wider).

To provide a measure of uncertainty on the model predictions and increase prediction accuracy we construct \gwsnmg \ as an ensemble of neural networks, utilizing the method of bagging in ensemble learning to train multiple base models and then aggregating the predictions \citep{Breiman1996}. We train 20 iterations of the same model, but each time with a different random splitting of the $10^4$ events in the training set for training and validation (this is similar to performing K-fold cross-validation, but here we allow the validation sets to have overlap). We then calculate and report the mean and standard deviation of the 20 model predictions, as the final \gwsnmg \ predicted probability and ensemble uncertainty.

\section{Results} \label{sec:results}

\subsection{Model performance on test data set}

\begin{figure}
\includegraphics[width=\columnwidth]{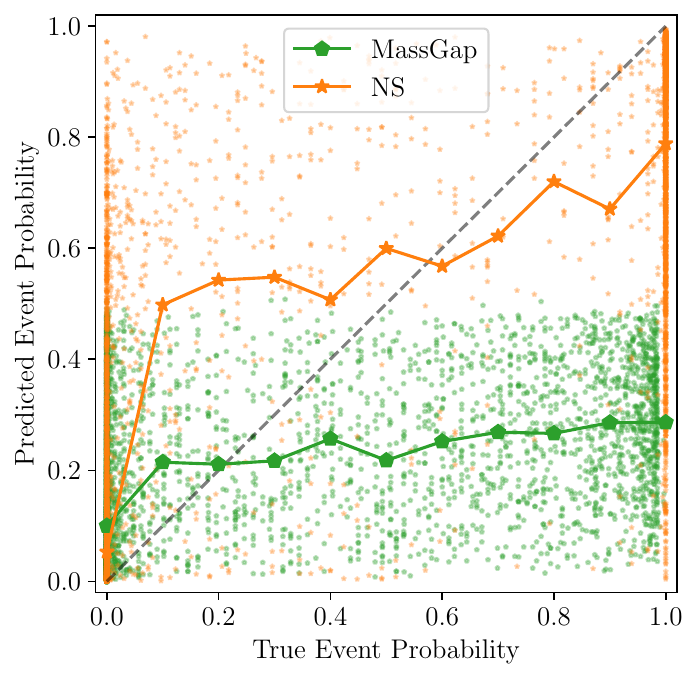}
\caption{Comparison of the predicted MassGap (green) and NS (orange) probabilities to the true probabilities for all $10^4$ events in our test set. Points that lie along the dashed gray diagonal line indicate perfectly accurate predictions. Overlaid are the mean values of the predictions binned by true probability. The majority of events in the test set have zero (or near-zero) true MassGap and NS probabilities, and so are clustered along the left-edge of the plot, where the binned mean predicted probability is also low ($<0.1$). The model is generally not able to accurately predict the true MassGap probabilities for non-zero events and is biased towards smaller predictions values. The NS predictions show a marginal increasing trend with true probability, but are also generally not accurate for non-extremal events. The mean absolute error for all $10^4$ events in the test set is 17\% for MassGap predictions and 10\% for NS predictions.
\label{fig:test_set_predicted_vs_true_probs}}
\end{figure}

We evaluate the \gwsnmg \ model on the hold-out test set of $10^4$ events to quantify how well it performs and to identify any prediction trends. Figure ~\ref{fig:test_set_predicted_vs_true_probs} shows how the model predicted MassGap and NS probabilities compare to the true values. To find general prediction trends we bin the predictions by their true probabilities and calculate the mean prediction probability for each bin. The model does relatively well in identifying events with zero (or near-zero) probability of MassGap and NS, which form the majority of events in the test set (as in reality): the mean absolute error for predictions in the lowest bin ($0 \leq p \leq 0.05$) is 10\% for MassGap and 5\% for NS. However, it struggles to predict the correct probability for events with larger, non-zero values ($p > 0.05$). The transition from low to high probabilities shows a marginal increasing trend for the NS predictions, but is virtually consistent with a uniform mean probability for the MassGap predictions of $\sim 0.2$. Furthermore, the MassGap output does not predict a value $\gtrsim 0.5$ for any event, showing a heavy bias for identifying events with low MassGap probability. Overall, the mean absolute error for all $10^4$ events in the test set is 17\% for MassGap predictions and 10\% for NS predictions. We aim to identify the source of this error in the predictions for marginal events by quantifying the model predictions as a function of the event source parameters.

The chirp mass $\mathcal{M}_c = (m_1m_2)^{3/5} / (m_1+m_2)^{1/5}$ is the combination of component masses that determines to leading (Newtonian) order both the amplitude of a GW signal at fixed orbital frequency and its frequency evolution, and is one of the most accurately measured source parameters for interferometric GW detectors (see, e.g., \cite{Cutler1993,Finn1993}). Along with the mass ratio $q = m_1/m_2$, these two quantities uniquely determine the component masses of the source (and thus the MassGap and NS probabilities).

\begin{figure*}
\includegraphics[width=\textwidth]{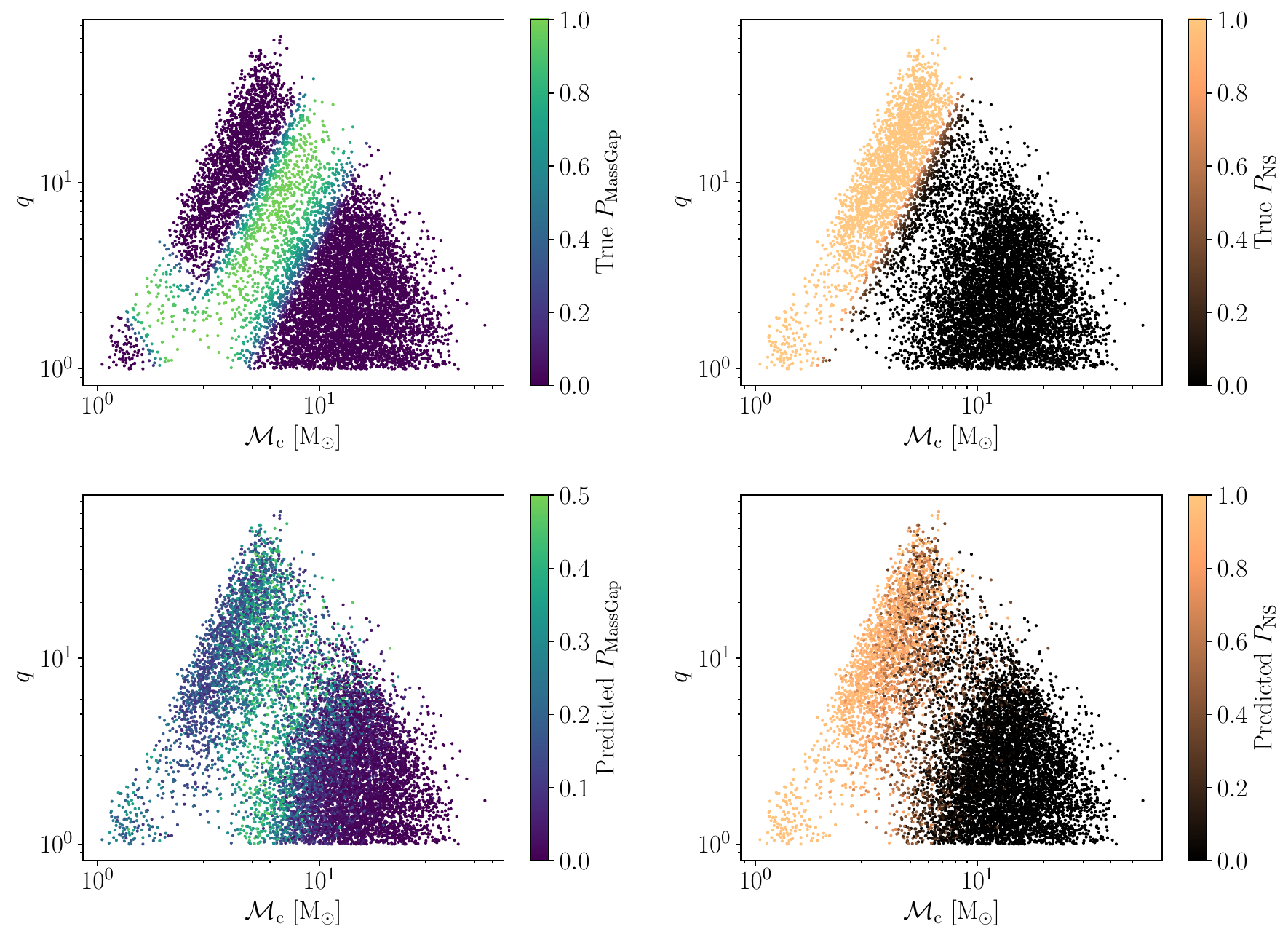}
\caption{Scatter plot showing the chirp mass ($\mathcal{M}_c$) and mass ratio ($q=m_1/m_2$) of the $10^4$ events in our test data set, and their true (top row) and \gwsnmg \ predicted (bottom row) MassGap and NS probabilities. The top row is comparable to Figure~\ref{fig:m1_m2_massgap_ns_prob}, which shows the true probabilities for these events as a function of the component masses. The model is able to learn broad trends for predicting the probability of the source based on the chirp mass, accurately predicting the near-zero MassGap and NS probabilities for high chirp mass events, and the high probability of NS for events with small chirp masses. However, it is unable to resolve the mass ratio dependence of the MassGap and NS probabilities for a given value of the chirp mass, leading to mixed predictions for intermediate chirp mass values that can support a wide range of true $P_{\mathrm{MassGap}}$ and $P_{\mathrm{NS}}$. Note that the MassGap predictions are heavily biased towards smaller probabilities and the maximum predicted $P_{\mathrm{MassGap}}$ is approximately 0.5 (indicated by the bottom left panel's color-bar range) --- the model is able to correctly predict sources that are likely not in the mass gap, but fails to identify the true mass gap sources with a high probability.
\label{fig:test_set_true_and_predicted_probs_vs_chirpmass_massratio}}
\end{figure*}

Figure~\ref{fig:test_set_true_and_predicted_probs_vs_chirpmass_massratio} shows the \gwsnmg \ predicted probabilities and true probabilities of the test set events distributed by their chirp mass and mass ratio. Comparing the true distribution (top row) to the predicted distribution (bottom row) we see that the model is able to learn broad features for MassGap and NS dependence in the $\mathcal{M}_c$ versus $q$ space. In particular, we see a strong dependence for the predicted probabilities on the chirp mass, and no (or marginally weak) dependence on the mass ratio. The model accurately predicts the near-zero MassGap and NS probabilities for high chirp mass events, but it's predictions are less accurate in the lower chirp mass regions where the value of the chirp mass does not uniquely determine the probability of MassGap or NS. That is, we find that the model uses the \textsc{Bayestar} localization inputs to infer information about the source chirp mass and then uses this information to predict the probability that the source has a mass gap or NS component.

In regions where different mass ratio values allow for a range of MassGap and NS probabilities, \gwsnmg \ predictions are not as accurate. This reflects the underlying degeneracy of the component masses for a known chirp mass. For example, a binary source with a chirp mass of $\mathcal{M}_c = 6 ~M_{\odot}$ can have component masses $m_1 = 6.9~M_{\odot}$, $m_2 = 6.9~M_{\odot}$ $(q=1)$ for which $P_{\mathrm{MassGap}} = 0$ and $P_{\mathrm{NS}} = 0$, or it could have $m_1 = 16.3~M_{\odot}$, $m_2 = 3.3~M_{\odot}$ $(q=5)$ for which $P_{\mathrm{MassGap}} \simeq 1$ and $P_{\mathrm{NS}} = 0$, or even $m_1 = 30.9~M_{\odot}$, $m_2 = 2.06~M_{\odot}$ $(q=15)$ for which $P_{\mathrm{MassGap}} \simeq 0$ and $P_{\mathrm{NS}} \simeq 1$. In other words, for a fixed chirp mass the true nature of the source could be a low mass ratio BBH merger or a high mass ratio MassGap or NSBH merger, and the \gwsnmg \ model is not able to accurately tell these cases apart.

\begin{figure*}
\includegraphics[width=\textwidth]{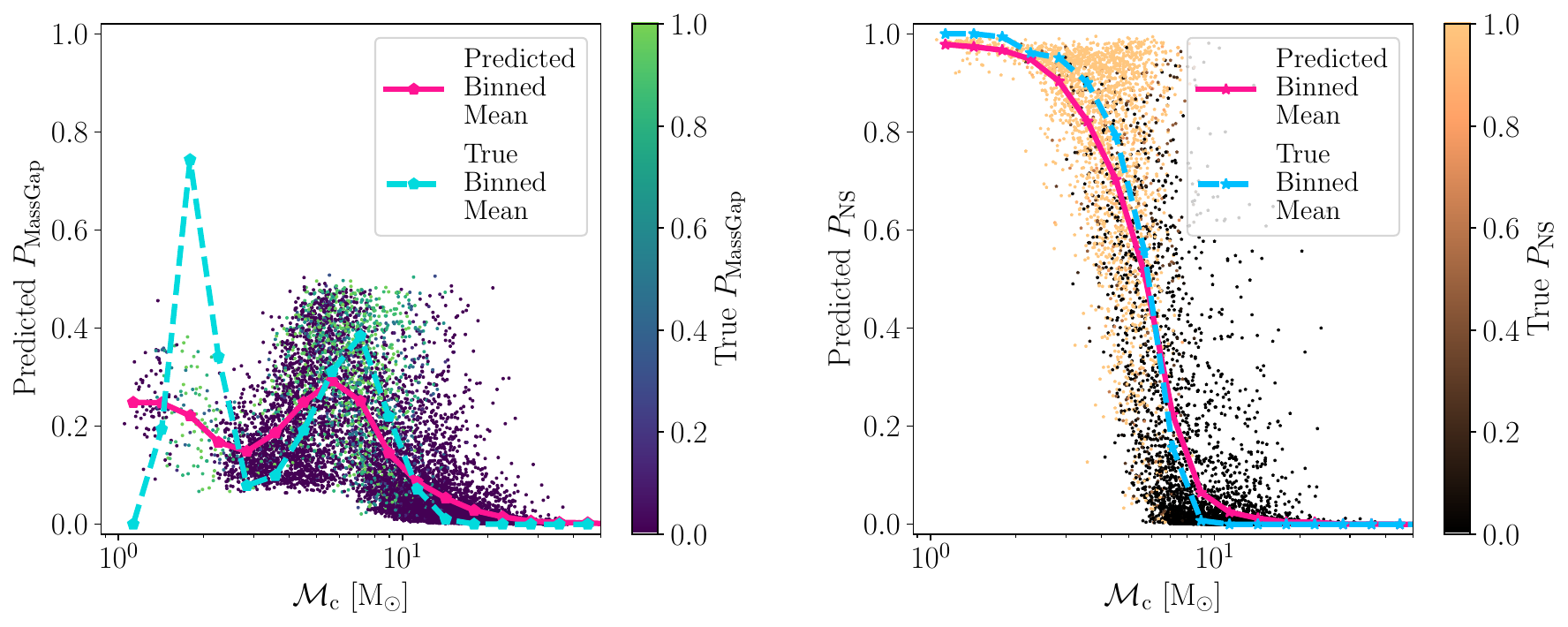}
\caption{Scatter plot showing the \gwsnmg \ predicted MassGap (left panel) and NS (right panel) probabilities of the $10^4$ events in our test set as a function of the source chirp mass. Each point is colored by its true MassGap and NS probability. Overlaid are the predicted (solid pink) and true (dashed blue) mean probabilities when the events are binned according to their chirp mass. For the NS predictions there is a strong monotonic chirp mass dependence, with the predicted mean probabilities closely following the true mean probabilities. The model is unable to accurately predict $P_{\mathrm{NS}}$ for events in the ambiguous range of $3~M_{\odot} \lesssim \mathcal{M}_c \lesssim 10~M_{\odot}$. For the MassGap predictions the chirp mass dependence is less evident, as the underlying true distribution is more complicated (multiple peaks), but the comparison of the predicted mean curve to the true mean curve shows that the model is still broadly learning the main chirp mass dependence features: near-zero probability for $\mathcal{M}_c \gtrsim 15~M_{\odot}$, higher probability for events around $\mathcal{M}_c \sim 7~M_{\odot}$, and lower probability for event with $\mathcal{M}_c \sim 3~M_{\odot}$. Overall the model is unable to discriminate sources to accurately predict $P_{\mathrm{MassGap}}$ for events with $\mathcal{M}_c \lesssim 15~M_{\odot}$.
\label{fig:test_set_predicted_probs_vs_chirp_mass}}
\end{figure*}

In Figure~\ref{fig:test_set_predicted_probs_vs_chirp_mass} we explicitly show this chirp mass dependence for the test event predictions. For the NS predictions there is a strong monotonic chirp mass dependence, with the predicted mean probabilities closely following the true mean probabilities. The model is not able to accurately predict $P_{\mathrm{NS}}$ for events in the ambiguous range of $3~M_{\odot} \lesssim \mathcal{M}_c \lesssim 10~M_{\odot}$. For the MassGap predictions the chirp mass dependence is less evident, as the underlying true distribution is more complicated (multiple peaks), but the model is still broadly learning the main chirp mass dependence features: near-zero probability for $\mathcal{M}_c \gtrsim 15~M_{\odot}$, higher probability for events around $\mathcal{M}_c \sim 7~M_{\odot}$, and lower probability for event with $\mathcal{M}_c \sim 3~M_{\odot}$. Overall the model is not able to accurately predict $P_{\mathrm{MassGap}}$ for events with $\mathcal{M}_c \lesssim 15~M_{\odot}$.

\subsection{Predictions and performance on O4a events}

The LVK collaboration has recently released its final catalog of candidate merger events for the first part of the fourth observing run (O4a), GWTC-4.0 \citep{Abac2025_gwtc4}. This catalog contains 128 detected events that are determined to have a probability of astrophysical CBC origin $p_{\mathrm{astro}} > 0.5$ in at least one of the search pipelines and are not vetoed during event validation. Of these, the LVK analyze the properties of a higher purity subset of 86 candidates with $\mathrm{FAR} < 1 ~yr^{-1}$, and provide the full parameter estimation data \citep{Abac2025_gwtc4_pedata}. 69 of these 86 are multi-detector (Hanford+Livingston) events that had an associated LVK public alert issued for a significant candidate with a \textsc{Bayestar} sky localization map. This subset of 69 events represents those events for which we have both the \gwsnmg \ prediction and the parameter estimation data from GWTC-4.0, which we can use to calculate the true MassGap and NS probabilities and compare to the model's predictions.

To calculate the GWTC-4.0 MassGap and NS probabilities we use the parameter estimation data release in \cite{Abac2025_gwtc4_pedata} along with Eqs.~\ref{eq:1}-\ref{eq:3}, marginalizing over all the hyperposterior samples for the component masses $m_1$ and $m_2$. We treat these values as the \textit{true} $P_{\mathrm{MassGap}}$ and $P_{\mathrm{NS}}$. We compare them to the \textit{predicted} \gwsnmg \ $P_{\mathrm{MassGap}}$ and $P_{\mathrm{NS}}$, for which we use the low-latency \textsc{Bayestar} localization maps released in the public alerts at the time of the events, as available on GraceDB\footnote{\url{https://gracedb.ligo.org/superevents/public/O4/}}. For events that were found by multiple search pipelines (superevents with multiple associated events), the predictions (and comparisons) shown are for the final preferred events as identified in low-latency by the LVK\footnote{\url{https://emfollow.docs.ligo.org/userguide/analysis/superevents.html}}.

We determine the overall performance of the model by calculating the mean absolute error of the \gwsnmg \ predicted probabilities to the GWTC-4.0 determined probabilities, and find that the mean error is 9\% for $P_{\mathrm{MassGap}}$ and 6\% for $P_{\mathrm{NS}}$. While these overall predicted errors are low, they are largely driven by the fact that the majority of events are correctly predicted to have a MassGap and NS probability of zero (or near zero). If we only consider the 18 events for which the \gwsnmg \ predicted probabilities are $> 0.1$, then the mean error increases to 31\% for $P_{\mathrm{MassGap}}$ and 23\% for $P_{\mathrm{NS}}$. For completeness, the \gwsnmg \ predicted probabilities as well as the GWTC-4.0 true probabilities for these 18 events are shown in Table~\ref{table:O4a_predictions}. The model correctly predicts $P_{\mathrm{NS}} = 0.98$ for the NSBH event GW2301518\_125908, and $P_{\mathrm{MassGap}} = 0.17$ for the lowest-mass BBH in O4a, GW230627\_015337. For the rest of the 16 events listed in the table, the \gwsnmg \ predictions do not match the GWTC-4.0 values.

\begin{table*}
\centering
\hspace*{-3.0cm}
\begin{tabular}{|l|c|c|c|c|c|c|}
\toprule
\toprule
\multicolumn{1}{|c|}{\textbf{Event ID}} & \multicolumn{1}{c|}{\textbf{GW Name}} & \multicolumn{3}{c|}{\textbf{GWTC-4.0 Properties}} & \multicolumn{2}{c|}{\textbf{\texttt{GWSkyNet-MassGap} Predictions}} \\
 &  & $\mathcal{M}_c ~[M_{\odot}]$ & $P_{\mathrm{MassGap}}$ & $P_{\mathrm{NS}}$ & $P_{\mathrm{MassGap}}$ & $P_{\mathrm{NS}}$ \\
\midrule
\textbf{S230518h} & \textbf{GW230518\_125908} & $2.8$ & 0.00 & $\mathbf{1.00}$ & $0.31 \pm 0.07$ & $\mathbf{0.98 \pm 0.01}$ \\
S230605o & GW230605\_065343 & $11.9$ & 0.00 & 0.00 & $0.45 \pm 0.02$ & $0.42 \pm 0.06$ \\
\textbf{S230627c} & \textbf{GW230627\_015337} & $6.02$ & $\mathbf{0.17}$ & 0.00 & $\mathbf{0.17 \pm 0.07}$ & $0.18 \pm 0.14$ \\
$\mathrm{S230630bq}^{\dagger}$ & GW230630\_234532 & $7.06$ & 0.07 & 0.00 & $0.35 \pm 0.03$ & $0.60 \pm 0.04 ^{\dagger}$ \\
S230706ah & GW230706\_104333 & $11.8$ & 0.00 & 0.00 & $0.14 \pm 0.04$ & $0.01 \pm 0.00$ \\
S230723ac & GW230723\_101834 & $11.4$ & 0.00 & 0.00 & $0.28 \pm 0.06$ & $0.06 \pm 0.02$ \\
S230729z & GW230729\_082317 & $8.33$ & 0.06 & 0.00 & $0.44 \pm 0.05$ & $0.24 \pm 0.07$ \\
S230731an & GW230731\_215307 & $7.77$ & 0.01 & 0.00 & $0.42 \pm 0.03$ & $0.33 \pm 0.08$ \\
S230904n & GW230904\_051013 & $7.54$ & 0.05 & 0.00 & $0.46 \pm 0.03$ & $0.43 \pm 0.05$ \\
$\mathrm{S230922q}^{\dagger *}$ & $\mathrm{GW230922\_040658}^*$ & $52^*$ & 0.00 & 0.00 & $0.37 \pm 0.07$ & $0.58 \pm 0.08^{\dagger}$ \\
$\mathrm{S231001aq}^*$ & $\mathrm{GW231001\_140220}^*$ & $46.6^*$ & 0.00 & 0.00 & $0.47 \pm 0.04$ & $0.28 \pm 0.06$ \\
S231020ba & GW231020\_142947 & $8.06$ & 0.12 & 0.00 & $0.26 \pm 0.03$ & $0.06 \pm 0.03$ \\
S231104ac & GW231104\_133418 & $8.84$ & 0.01 & 0.00 & $0.33 \pm 0.05$ & $0.10 \pm 0.04$ \\
S231113bw & GW231113\_200417 & $8.01$ & 0.04 & 0.00 & $0.44 \pm 0.04$ & $0.25 \pm 0.06$ \\
S231114n & GW231114\_043211 & $11.6$ & 0.01 & 0.00 & $0.16 \pm 0.04$ & $0.01 \pm 0.00$ \\
S231118an & GW231118\_090602 & $8.37$ & 0.13 & 0.00 & $0.41 \pm 0.05$ & $0.20 \pm 0.06$ \\
$\mathrm{S231223j}^*$ & $\mathrm{GW231223\_032836}^*$ & $31.8^*$ & 0.00 & 0.00 & $0.31 \pm 0.04$ & $0.09 \pm 0.03$ \\
S231224e & GW231224\_024321 & $7.13$ & 0.01 & 0.00 & $0.40 \pm 0.04$ & $0.24 \pm 0.07$ \\
\bottomrule
\end{tabular}
\caption{Predictions for O4a events for which the \gwsnmg \ predictions have $P_{\mathrm{MassGap}} > 0.1$ or $P_{\mathrm{NS}} > 0.1$. For each of these 18 events, we also show the corresponding event properties derived from GWTC-4.0. For the chirp mass ($\mathcal{M}_c$) we quote the median value as stated in \cite{Abac2025_gwtc4}. For calculating the GWTC-4.0 MassGap and NS probabilities we use the parameter estimation data release in \cite{Abac2025_gwtc4_pedata} along with Eqs.~\ref{eq:1}-\ref{eq:3}, marginalizing over all the hyperposterior samples for the component masses $m_1$ and $m_2$. For the rest of the 51 events, the \gwsnmg \ predictions have $P_{\mathrm{MassGap}} < 0.1$ and $P_{\mathrm{NS}} < 0.1$, and all the GWTC-4.0 determined probabilities are zero. We highlight in \textbf{bold} the event S230518h/GW230518\_125908 for which the predicted \gwsnmg \ $P_{\mathrm{NS}}$ closely matches the true GWTC-4.0 $P_{\mathrm{NS}}$, and the event S230627c/GW230627\_015337 for which the predicted \gwsnmg \ $P_{\mathrm{MassGap}}$ closely matches the true GWTC-4.0 $P_{\mathrm{MassGap}}$. For the rest of the 16 events listed in the table, the \gwsnmg \ predictions do not match the GWTC-4.0 values; two of these events have model predictions with high probability $P > 0.5$, highlighted with a dagger ($\dagger$). All of these events have $\mathcal{M}_c < 15 ~M_{\odot}$, except for three events that are highlighted with an asterisk ($*$); these three events are also outliers in terms of how significantly their distances were under-estimated in the low-latency \textsc{Bayestar} localization (see Figure~\ref{fig:O4a_median_distance_vs_input_distance}).}
\label{table:O4a_predictions}
\end{table*}

\begin{figure}
\includegraphics[width=\columnwidth]{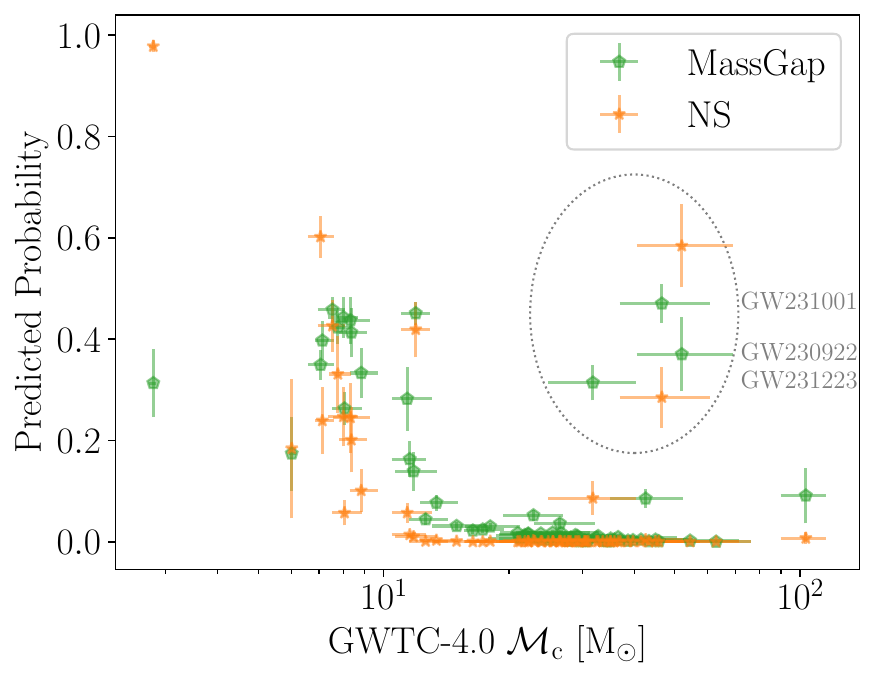}
\caption{Model predicted MassGap and NS probabilities as a function of the source chirp mass (as determined in GWTC-4.0) for the 69 events in O4a analyzed in this work. Similar trends as seen for the test data predictions confirm the model's learned chirp mass dependence: significant predicted probabilities for events with $\mathcal{M}_c < 15 ~M_{\odot}$, and zero (or near-zero) otherwise. The three outlier events that have large chirp masses but are still predicted to have $P_{\mathrm{MassGap}} > 0.1$ or $P_{\mathrm{NS}} > 0.1$ are GW230922\_040658, GW231001\_140220, and GW231223\_032836 (encircled). The online low-latency analysis underestimates the distance for these events, leading \gwsnmg \ to likely infer these events to have much smaller chirp masses and thus inaccurately predict a higher probability of being a MassGap (or NS) event.
\label{fig:O4a_predicted_probs_vs_chirp_mass}}
\end{figure}

Figure~\ref{fig:O4a_predicted_probs_vs_chirp_mass} shows how the \gwsnmg \ predicted probabilities for all 69 events in O4a analyzed in this work vary according to their determined chirp mass values in GWTC-4.0 \citep{Abac2025_gwtc4}. The same trends in chirp mass dependence for the predicted probabilities appear for these O4a events as was seen for the test data set: the predicted probabilities are zero or near-zero for events with $\mathcal{M}_c > 15 ~M_{\odot}$ (aside from three outlier events, discussed below), and increase to larger values for the lower-mass events, confirming our conclusions from the test data that the model is learning to infer information about the source chirp mass to make its $P_{\mathrm{MassGap}}$ and $P_{\mathrm{NS}}$ predictions.

\begin{figure}
\includegraphics[width=\columnwidth]{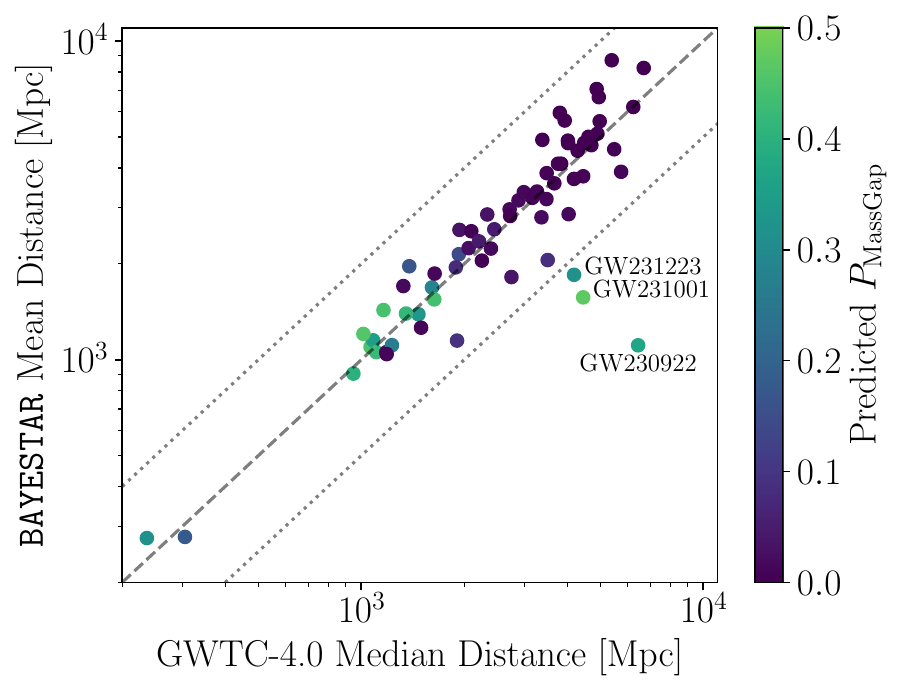}
\caption{Comparison of the final determined value of the distance to the source (GWTC-4.0 median distance) versus the online low-latency determined value (\textsc{Bayestar} mean distance) for the 69 events in O4a analyzed in this work. The points are colored according to the MassGap event probability predicted by \gwsnmg. The dashed gray line along the diagonal indicates where the \textsc{Bayestar} distance is equal to the GWTC-4.0 distance, $d_{\mathrm{BAYESTAR}} = d_{\mathrm{GWTC-4.0}}$. The dotted gray lines above and below it indicate points where $d_{\mathrm{BAYESTAR}} = 2\times d_{\mathrm{GWTC-4.0}}$ (above) and $d_{\mathrm{BAYESTAR}} = 0.5\times d_{\mathrm{GWTC-4.0}}$ (below). While most points lie close to the diagonal line and within a factor of two of each other (bounded by the dotted gray lines), indicating good agreement between the final distance and the low-latency determined value, there are three events that are clear outliers, indicated on the figure with their shortened GW candidate event name. Since \gwsnmg \ takes the \textsc{Bayestar} mean distance value as one of its inputs, the large under-estimation of the distance as compared to the true value influences its determination of the source properties: the model infers these events to have much smaller chirp masses and thus predicts a higher probability of being a MassGap (or NS) event, as seen in Figure~\ref{fig:O4a_predicted_probs_vs_chirp_mass}.
\label{fig:O4a_median_distance_vs_input_distance}}
\end{figure}

There are three events that are clear outliers in this analysis --- events that have large chirp masses $\mathcal{M}_c > 15 ~M_{\odot}$ but are still predicted to have $P_{\mathrm{MassGap}} > 0.1$ or $P_{\mathrm{NS}} > 0.1$. These are events GW230922\_040658, GW231001\_140220, and GW231223\_032836. Figure~\ref{fig:O4a_median_distance_vs_input_distance} shows that these three events are also outliers in terms of how closely their low-latency estimate of mean distance from the online \textsc{Bayestar} localization matches the final determined distance as included in GWTC-4.0. The online analysis underestimates the distance by more than a factor of two for these events, leading \gwsnmg \ to infer that these events have much smaller chirp masses and thus inaccurately predict a higher probability of being a MassGap (or NS) event. For GW231001\_140220 and GW231223\_032836, this low-latency under-estimation is likely caused by the presence and influence of noise artifacts around the time of the event, for which glitch mitigation was required in the offline analysis before full parameter estimation could be performed \citep{Abac2025_gwtc4,Abac2025_gwtc4_methods}.

There are two events included in GWTC-4.0 that have $P_{\mathrm{NS}} > 0.1$: GW230518\_125908 and GW230529\_181500. We do not predict on the event GW230529 or include it in our O4a analysis because it was a single detector event, which \gwsnmg \ is not trained to predict on. However, as this was a high-interest NSBH event, we simulate a GW230529-like event 1000 times by fixing the component masses to be the median determined values for the event in GWTC-4.0, $m_1 = 3.66 ~M_{\odot}$ and $m_2 = 1.42 ~M_{\odot}$ \citep{Abac2025_gwtc4}, and then draw random samples for all other source and extrinsic parameters using the same scheme we used for generating our astrophysical merger events as outlined in Section~\ref{sec:methods}.1. We then calculate the predicted probabilities for all 1000 of the simulated GW230529-like events, and aggregate the results to find that the \gwsnmg \ prediction is $P_{\mathrm{MassGap}} = 0.21^{+0.12}_{-0.12}$ and $P_{\mathrm{NS}} = 0.96^{+0.02}_{-0.05}$ (median and 90\% credible intervals). The true probabilities for GW230529, as determined from its source mass estimates in GWTC-4.0, are $P_{\mathrm{MassGap}} = 0.90$ and $P_{\mathrm{NS}} = 0.99$. Thus the \gwsnmg \ predicted $P_{\mathrm{NS}}$ is close to the true value but $P_{\mathrm{MassGap}}$ is underestimated. The median chirp mass for the event as determined in GWTC-4.0 is $\mathcal{M}_c = 1.94 ~M_{\odot}$; the model mean predicted probabilities at a chirp mass of $1.94 ~M_{\odot}$ are $P_{\mathrm{MassGap}} \simeq 0.20$ and $P_{\mathrm{NS}} \simeq 0.95$ (Figure~\ref{fig:test_set_predicted_probs_vs_chirp_mass}, pink solid lines), closely matching the GW230529-like event prediction values. That is, we once again find that the model prediction is largely determined by the inferred source chirp mass. This is further supported by the fact that the predicted $P_{\mathrm{MassGap}}$ and $P_{\mathrm{NS}}$ for the GW230529-like events span a small range of values, even though only the mass is fixed for these 1000 simulated events while all other parameters vary considerably and cover the entire detectable range (e.g., the source distance varies between $\sim 100-800 ~\mathrm{Mpc}$).

\section{Model and data set limitations}\label{sec:model_data_limitations}

For NSs with non-zero spin, the maximum mass of a NS can be higher than the non-rotating TOV mass due to additional rotational support. For a NS rotating uniformly at its mass-shedding (Keplerian) limit, the maximum supported mass can be up to $M_{\mathrm{NS,max}} \simeq 1.203 ~M_{\mathrm{NS,TOV}}$ \citep[a universal relation irrespective of the NS equation of state considered;][]{Breu2016}. However, since the PDB model that we use to generate our data does not model the NS (or BH) spin and assumes a uniform distribution, we do not account for the effects of spin in this work. Furthermore, using the empirical fit relations described in \cite{Breu2016,Most2020}, we calculate that even for high spin NSs with $\chi = 0.4$ (the maximum value assumed for objects with mass $m < 2.5 M_{\odot}$ in the PDB model), the spin-corrected maximum NS mass is $M_{\mathrm{NS,max}} \simeq 1.06 ~M_{\mathrm{NS,TOV}}$, a marginal increase of 6\%. In future work we will explore sampling from a more astrophysically-constrained distribution of component spins, and then take into account the effects of spin on the maximum allowed NS mass for these samples.

We have also presented the results of our model for which we calculated the NS and MassGap probabilities based on the maximum TOV mass \citep{Legred2021} and the minimum BH mass from the \textsc{Power Law + Peak} (PP) model \citep{Abbott2023_gwtc3_population}. An alternate model can be trained on the MassGap probabilities based on the PDB distribution, which explicitly models the location of the lower and upper mass gap, $M^{\mathrm{gap}}_{\mathrm{low}}$ and $M^{\mathrm{gap}}_{\mathrm{high}}$. The advantage of this approach is that it allows for the location of the MassGap boundaries to be independent of the maximum NS and minimum BH mass. In practice, however, when the PDB model is fit to events included up to GWTC-3, the posterior distributions of both these parameters have broad support over the prior range: $M^{\mathrm{gap}}_{\mathrm{low}} = 2.1^{+0.8}_{-0.6}$ and $M^{\mathrm{gap}}_{\mathrm{high}} = 5.6^{+1.7}_{-1.8}$ (90\% credible intervals) \citep{Abbott2023_gwtc3_population}. Furthermore, as noted in \cite{Abbott2023_gwtc3_population}, the location and depth of the mass gap changes significantly depending on the mass pairing function assumed, and so impacts the classification of components in the mass gap. Recent work analyzing the effects of more flexible population models than PDB on the NS classification probability also confirms this dependence on the pairing function \citep{Mali2026}. Thus determining the MassGap probabilities based on the PDB parameters further blurs the boundaries of the MassGap, making it more difficult for the model to learn this boundary and determine the nature of the event. In our studies we also train a model based on the PDB boundary probabilities, and find that the results are qualitatively similar to training the model on the TOV+PP boundary probabilities, but with higher uncertainty and confusion between sources. We thus limit the results presented in this work to the TOV+PP model.

\section{Conclusion} \label{sec:conclusion}

GW signals from compact binary merger events detected by the LVK indicate that the purported lower mass gap ($\sim 2-5 ~M_{\odot}$) is likely not empty, and there are merger events with components that straddle the boundary between NSs and BHs. Distinguishing whether a candidate GW binary merger event has components in this mass gap can indicate the likelihood of it having a detectable electromagnetic counterpart, and thus inform decisions for follow-up observations. With the LVK having recently announced an interim six-month observing run between O4 and O5 (designated IR1), scheduled to begin in late 2026\footnote{\url{https://observing.docs.ligo.org/plan/}}, the next mass gap merger event could be detected soon.

In this work we have expanded upon the \gwsnm \ classifier and built a robust, astrophysically motivated set of simulated binary mergers to train a neural network model, \gwsnmg, that simultaneously predicts the probability that a candidate binary merger event has a component in the lower mass gap and the probability that it has a component which is a neutron star. These properties are also predicted by the LVK in low-latency using machine learning models and reported as part of the public alert contents (\texttt{HasMassGap} and \texttt{HasNS}). \gwsnmg \ differs in that the model is fully open and reproducible outside of the LVK collaboration, taking in only publicly available data at the time of the event. It thus serves as an alternate tool for the community to inform follow-up decisions.

Using information contained in the low-latency localization maps released in the LVK public alerts as model inputs, we find that the model is able to infer information about the source chirp mass to predict the MassGap and NS probabilities. This learned dependence leads to accurate predictions for high-mass mergers ($\mathcal{M}_c \gtrsim 15~M_{\odot}$), which have $P_{\mathrm{MassGap}} \sim 0$ and $P_{\mathrm{NS}} \sim 0$. However, the model predictions are not accurate for low-mass systems with $3~M_{\odot} \lesssim \mathcal{M}_c \lesssim 15~M_{\odot}$, corresponding to the range where the chirp mass does not uniquely determine the probability of MassGap or NS; without additional information about the binary mass ratio to break the degeneracy, the model can not accurately discriminate between sources that have similar $\mathcal{M}_c$ but different $P_{\mathrm{MassGap}}$ and $P_{\mathrm{NS}}$. For candidate events in the first part of the LVK fourth observing run (O4a), the model has a mean prediction error of 9\% for $P_{\mathrm{MassGap}}$ and 6\% for $P_{\mathrm{NS}}$. Our work points to the possibility that the model could be further developed to rapidly predict the source chirp mass for candidate events in future observing runs.

We provide the \gwsnmg \ model predictions for all O4 candidate events on a publicly accessible webpage\footnote{\url{https://nayyer-raza.github.io/projects/GWSkyNet-MassGap/}}, which will be updated with real-time predictions as candidate events are detected in future observing runs IR1 and O5. We also provide the model and associated scripts in a git repository\footnote{\url{https://github.com/nayyer-raza/GWSkyNet-Multi}} for users who wish to implement it locally in their own follow-up pipelines.

\begin{acknowledgments}
The authors acknowledge support for this project from the Canadian Tri-Agency New Frontiers in Research Fund - Exploration program, and from the Canadian Institute for Advanced Research (CIFAR), in particular the CIFAR Catalyst program in supporting the ML-ESTEEM collaboration. N.R. is supported by a Walter C. Sumner Memorial Fellowship and acknowledges funding support from the Trottier Space Institute at McGill. D.H. and J.M. acknowledge support from the Natural Sciences and Engineering Research Council of Canada (NSERC) Discovery Grant program and the Canada Research Chairs (CRC) program. D.H. acknowledges support from the NSERC Arthur B. McDonald Fellowship. A.M. acknowledges support from the NSF (1640818, AST-1815034). A.M. and J.M. also acknowledge support from IUSSTF (JC-001/2017). This material is based upon work supported by NSF’s LIGO Laboratory, which is a major facility fully funded by the National Science Foundation.
\end{acknowledgments}



\appendix

\section{Predictions for events in O4b and O4c}
There are 186 significant alerts issued by the LVK for multi-detector events with \textsc{Bayestar} sky localization maps in O4b and O4c. In Table~\ref{table:O4b_O4c_predictions} we provide the \gwsnmg \ predictions for the subset of 66 events that have not been retracted and have predicted probabilities $P_{\mathrm{MassGap}} > 0.1$ or $P_{\mathrm{NS}} > 0.1$.

\begin{table*}
\centering
\hspace*{-3.0cm}
\begin{tabular}{|l|c|c|c|l|c|c|}
\toprule
\toprule
\multicolumn{3}{|c|}{\textbf{O4b}} & ~~~~~~ & \multicolumn{3}{|c|}{\textbf{O4c}} \\
\midrule

\multicolumn{1}{|c|}{\textbf{Event ID}} & \multicolumn{2}{c|}{\textbf{\texttt{GWSkyNet-MassGap} Predictions}} &  & \multicolumn{1}{c|}{\textbf{Event ID}} & \multicolumn{2}{c|}{\textbf{\texttt{GWSkyNet-MassGap} Predictions}} \\
 & $P_{\mathrm{MassGap}}$ & $P_{\mathrm{NS}}$ & & & $P_{\mathrm{MassGap}}$ & $P_{\mathrm{NS}}$ \\
\midrule
S240413p & $0.29 \pm 0.04$ & $0.43 \pm 0.07$ & & S250205bk & $0.28 \pm 0.04$ & $0.43 \pm 0.10$ \\
\textit{S240422ed} & $\mathit{0.20 \pm 0.08}$ & $\mathit{0.99 \pm 0.01}$ & & S250206dm & $0.32 \pm 0.06$ & $0.97 \pm 0.01$ \\
S240426s & $0.23 \pm 0.07$ & $0.98 \pm 0.01$ & & S250211aa & $0.24 \pm 0.02$ & $0.06 \pm 0.02$ \\
S240428dr & $0.22 \pm 0.05$ & $0.03 \pm 0.01$ & & S250304cb & $0.39 \pm 0.03$ & $0.14 \pm 0.04$ \\
S240507p & $0.43 \pm 0.03$ & $0.34 \pm 0.05$ & & S250328ae & $0.22 \pm 0.07$ & $0.11 \pm 0.05$ \\
S240512r & $0.37 \pm 0.03$ & $0.47 \pm 0.04$ & & S250331o & $0.18 \pm 0.04$ & $0.05 \pm 0.03$ \\
S240520cv & $0.32 \pm 0.03$ & $0.21 \pm 0.06$ & & S250628am & $0.33 \pm 0.04$ & $0.25 \pm 0.07$ \\
S240527fv & $0.17 \pm 0.03$ & $0.04 \pm 0.01$ & & S250629bs & $0.17 \pm 0.03$ & $0.02 \pm 0.01$ \\
S240530a & $0.41 \pm 0.03$ & $0.27 \pm 0.05$ & & S250704ab & $0.24 \pm 0.04$ & $0.72 \pm 0.05$ \\
S240601co & $0.41 \pm 0.03$ & $0.47 \pm 0.04$ & & S250705cb & $0.38 \pm 0.03$ & $0.30 \pm 0.04$ \\
S240627by & $0.42 \pm 0.03$ & $0.39 \pm 0.05$ & & S250725j & $0.27 \pm 0.09$ & $0.71 \pm 0.09$ \\
S240629by & $0.21 \pm 0.03$ & $0.06 \pm 0.02$ & & S250725l & $0.12 \pm 0.02$ & $0.01 \pm 0.01$ \\
S240807h & $0.43 \pm 0.04$ & $0.42 \pm 0.05$ & & S250727dc & $0.12 \pm 0.03$ & $0.02 \pm 0.01$ \\
S240813c & $0.34 \pm 0.05$ & $0.18 \pm 0.06$ & & S250810ck & $0.14 \pm 0.04$ & $0.85 \pm 0.05$ \\
S240825ar & $0.40 \pm 0.03$ & $0.51 \pm 0.04$ & & \textit{S250818k} & $\mathit{0.18 \pm 0.08}$ & $\mathit{0.99 \pm 0.01}$ \\
S240830gn & $0.41 \pm 0.03$ & $0.40 \pm 0.05$ & & S250818t & $0.26 \pm 0.03$ & $0.08 \pm 0.02$ \\
S240910ci & $0.46 \pm 0.06$ & $0.27 \pm 0.06$ & & S250827fo & $0.30 \pm 0.03$ & $0.66 \pm 0.04$ \\
S240915b & $0.21 \pm 0.04$ & $0.07 \pm 0.03$ & & S250830bp & $0.28 \pm 0.04$ & $0.26 \pm 0.06$ \\
S240915bd & $0.29 \pm 0.09$ & $0.64 \pm 0.12$ & & S250901cb & $0.11 \pm 0.01$ & $0.01 \pm 0.00$ \\
S240916ar & $0.44 \pm 0.03$ & $0.36 \pm 0.05$ & & S250906ca & $0.38 \pm 0.04$ & $0.40 \pm 0.05$ \\
S240921cw & $0.11 \pm 0.03$ & $0.01 \pm 0.01$ & & S250908y & $0.36 \pm 0.05$ & $0.07 \pm 0.03$ \\
S240922df & $0.36 \pm 0.03$ & $0.30 \pm 0.04$ & & S250911ac & $0.42 \pm 0.04$ & $0.46 \pm 0.07$ \\
S240925n & $0.20 \pm 0.07$ & $0.71 \pm 0.13$ & & S250917aq & $0.34 \pm 0.11$ & $0.57 \pm 0.14$ \\
S240930aa & $0.10 \pm 0.02$ & $0.01 \pm 0.01$ & & S251006dd & $0.28 \pm 0.04$ & $0.66 \pm 0.04$ \\
S241009an & $0.36 \pm 0.04$ & $0.27 \pm 0.09$ & & S251013x & $0.19 \pm 0.03$ & $0.02 \pm 0.01$ \\
S241011k & $0.43 \pm 0.16$ & $0.37 \pm 0.12$ & & S251026bn & $0.40 \pm 0.04$ & $0.19 \pm 0.06$ \\
S241102br & $0.13 \pm 0.05$ & $0.87 \pm 0.06$ & & S251103f & $0.12 \pm 0.02$ & $0.01 \pm 0.00$ \\
S241109bn & $0.08 \pm 0.03$ & $0.94 \pm 0.03$ & & S251108dn & $0.39 \pm 0.04$ & $0.46 \pm 0.05$ \\
S241110br & $0.27 \pm 0.04$ & $0.70 \pm 0.05$ & & \textit{S251112cm} & $\mathit{0.15 \pm 0.09}$ & $\mathit{1.00 \pm 0.00}$ \\
S241114bi & $0.42 \pm 0.05$ & $0.51 \pm 0.07$ & &  &  &  \\
S241127aj & $0.11 \pm 0.03$ & $0.01 \pm 0.01$ & &  &  &  \\
S241130be & $0.38 \pm 0.03$ & $0.54 \pm 0.04$ & &  &  &  \\
S241225c & $0.45 \pm 0.08$ & $0.20 \pm 0.08$ & &  &  &  \\
S241231bg & $0.29 \pm 0.04$ & $0.16 \pm 0.08$ & &  &  &  \\
S250109bi & $0.21 \pm 0.05$ & $0.02 \pm 0.02$ & &  &  &  \\
S250118az & $0.43 \pm 0.03$ & $0.44 \pm 0.05$ & &  &  &  \\
S250119cv & $0.27 \pm 0.05$ & $0.16 \pm 0.06$ & &  &  &  \\
\bottomrule
\end{tabular}
\caption{The \gwsnmg \ predictions for the subset of candidate events in O4b and O4c that have not been retracted and have predicted probabilities $P_{\mathrm{MassGap}} > 0.1$ or $P_{\mathrm{NS}} > 0.1$. Events for which the \gwsnmm \ model predicts the source to likely be a glitch, S240422ed, S250818k, and S251112cm, are highlighted in \textit{italics}.}
\label{table:O4b_O4c_predictions}
\end{table*}

\bibliography{references}{}
\bibliographystyle{aasjournalv7}



\end{document}